\documentclass[aps,pra, twocolumn,superscriptaddress,amsmath,amssymb,floatfix]{revtex4}
\usepackage{graphicx}
\usepackage{times}
\usepackage{caption}
\usepackage{graphicx,amsmath,amssymb,amstext,amsfonts,color}

\usepackage{amsmath}
\usepackage{amssymb}
\def\D{\Delta}
\def\d{\delta}
\def\r{\rho}
\def\p{\pi}
\def\a{\alpha}
\def\g{\gamma}
\def\ra{\rightarrow}
\def\s{\sigma}
\def\b{\beta}
\def\e{\epsilon}
\def\G{\Gamma}
\def\om{\omega}
\def\l{\lambda}
\def\f{\phi}
\def\w{\psi}
\def\m{\mu}
\def\t{\tau}
\def\c{\chi}

\usepackage{times}
\DeclareMathOperator{\bx}{\mathbf{x}}

\begin{document}

\title{Efficient Simulations of Individual Based Models for Adaptive Dynamics and the Canonical Equation}
\author{Vaibhav Madhok}
\email{vmadhok@gmail.com}
\affiliation{Department of Zoology, University of British Columbia, 6270 University Boulevard, Vancouver B.C. Canada, V6T 1Z4}

\def\T{\Theta}
\def\D{\Delta}
\def\d{\delta}
\def\r{\rho}
\def\p{\pi}
\def\a{\alpha}
\def\g{\gamma}
\def\ra{\rightarrow}
\def\s{\sigma}
\def\b{\beta}
\def\e{\epsilon}
\def\G{\Gamma}
\def\om{\omega}
\def\l{\lambda}
\def\f{\phi}  
\def\w{\psi}
\def\m{\mu}
\def\t{\tau}
\def\c{\chi}
\begin{abstract}

%1) logistic map....any population size will work....in reconstructing the model
%2) when there is a borth rate as well as a death rate...one would expect departure from adaptive dynamics....as there are more births than deaths
%3) to modify that ....use a moran kind of model with your algorithm.

We propose a faster algorithm for individual based simulations for adaptive dynamics based on a simple modification to the standard Gillespie Algorithm for simulating stochastic birth death processes. We provide an analytical explanation that shows that simulations based on the modified algorithm, in the deterministic limit, lead to the same equations of adaptive dynamics as well as same conditions for evolutionary branching as those obtained from the standard Gillespie algorithm. 
 Based on this algorithm, we provide an intuitive and simple interpretation of 
the canonical equation of adaptive dynamics. 
With the help of examples we compare the performance of this algorithm to the standard Gillespie algorithm %\cite{Gillespie1976}
 and demonstrate its efficiency. We also study an example using this algorithm to study evolutionary dynamics in a multi-dimensional phenotypic space and study the question of predictability of evolution.
%We also discuss how the modified and the standard algorithms are two limits of a family of ecological models with a variable degree of competitive interaction and how this family of ecological models will lead to the same adaptive dynamics.
\end{abstract}

\maketitle

\vskip 0.5 cm
%begin{twocolumn}
\section{Introduction} 
 Evolution is essentially a birth death process involving mutations and selection. According to the traditional view, evolution optimizes scalar phenotypes like beak size, fecundity, body size etc.  The dictum, ``Survival of the fittest", leads to the fittest type ``winning". Evolution is thus viewed as a dynamical system converging to an equilibrium or a steady state. Such a view fails to take into account interactions among the individuals comprising a population or interactions among various individuals present in a geographical location.
 The birth and death rates of the individual depend on its phenotypic composition as well as its interaction with the ``environment". The ``environment" consists of both the ``abiotic" environment like temperature, climate etc. and the ``biotic" 
 environment which can be, for example, the types of all other individuals with which it interacts, for example, the competition it faces with regards to resources such as food and shelter. Another example of the ``biotic" environment is the presence of ``prey" or ``predator" type of individuals in the population \cite{mdbook}.
  Studying evolution involves modeling this birth death process in order to address questions about origin of diversity, speciation
 and predictability of evolution. 
In order to have such a view, we need to take into account certain ecological mechanisms like the interaction among organisms present in the same geographical location - a phenomenon known as frequency dependence \cite{mdbook, Diekmann04, DieckmannDoebeli99, Geritz98}. 
 Adaptive dynamics is a framework that incorporates frequency dependent selection originating from ecological interactions and competition among species \cite{mdbook}.  Under this framework, evolution is regarded as a continuous trajectory in the phenotypic space that unravels as rare mutants with a higher ``fitness" invade a resident population. The evolutionary dynamics in the above framework is described by
 the `` canonical equation" \cite{dieckmann96}, which, for one dimension, is given by
 \begin{align}
 \label{AD_basic}
 \frac{d{x}}{dt} = \gamma \frac{\partial f(\bold{x}, \bold{y})}{\partial y}_{\big | \bold{y}=\bold{x}}.
\end{align}  

 Here $\gamma$ is a constant of proportionality that scales the rate of evolution. The function, $ f(\bold{x}, \bold{y})$, is the measure of fitness of a rare mutant, $y$, in a monomorphic population with a resident trait $x$ \cite{dieckmann96}. %The function $f(\bold{x}, \bold{y})$ is the invasion fitness function that describes the per capita growth rate of a rare mutant, $y$, in the resident population at phenotype value $x$. 
The derivative of the invasion fitness function, $f(\bold{x}, \bold{y})$, with respect to the mutant phenotype coordinate, $y$, gives the direction of motion of the evolutionary trajectory. Mutations cause the occurrence of rare types which can possibly invade the resident depending on their relative growth rate with respect to the resident which is given by the selection gradient (RHS of Eq \ref{AD_basic}).

 The derivation of the canonical equation takes into account the stochastic nature of the birth-death process
and is based on the ecological processes influencing the birth and death at the level of an individual \cite{dieckmann96, champagnat08, champagnat06}. Such a formulation assumes that mutations are small, they occur very rarely and the resident population is taken to be of infinite size \cite{champagnat08}. 
The canonical equation (Eq. \ref{AD_basic}) describes the dynamics of a monomorphic population in the trait space driven by mutations, selection and invasion. 
 
In order to address questions like the origin of diversity, to study the stochastic effects in evolution and to study the issue of predictability of evolution we need to construct individual based models.
Individual based stochastic models based on the deterministic adaptive dynamics are a natural generalization of the latter. In these models, we represent the population as a ``cloud" of points in the phenotypic space. Each individual, represented by a point in the cloud, has a given birth and death rate
depending on the location of the point in the phenotypic space. Individuals with a higher birth rate have a greater probability to reproduce and the individuals with a higher death rate have a greater probability to die. 

Individual-based realizations of the model
are typically based on the Gillespie algorithm \cite{Gillespie1976}. Often, simulations of these models take a long time for one to see interesting effects like diversification in the phenotype space. Moreover, once diversification occurs, especially in high dimensional phenotypic spaces, it usually leads to an increase in
population of individuals and therefore making simulations even slower.
In this work, we accomplish three things.
First, we propose a simple modification 
of the standard Gillespie algorithm and make the individual based simulations more efficient.
With the help of numerical simulations, we demonstrate how this leads to faster simulations.
Secondly, using this algorithm and under certain assumptions, we give a very intuitive and simple interpretation of the canonical equation of adaptive dynamics. Thirdly, we discuss the issue of the predictability of evolution by considering individuals based models in high dimensional phenotypic spaces and demonstrate the ``evolutionary butterfly effect".

 %The remainder of the paper is organized as follows: Our starting point is individual-based formulation of general competition models describing ecological interactions that occur in logistic type equations. We then briefly discuss the adaptive dynamics limit of this formulation. We then describe the standard individual based simulation and give a simple modification to the algorithm. After demonstrating the advantage of the modification, we derive the canonical equation of adaptive dynamics from it. 

 \section{Individual based models to macroscopic deterministic adaptive dynamics}
 
We begin with the widely studied logistic model \cite{may}, 
 
 \begin{align}
\label{logistic}
 \frac{\partial N(\bold{x}, t)}{\partial t} = r N(\bold{x}, t)\bigg( 1 - \frac{\int \alpha(\bold{x}, \bold{y}) N (\bold{y}, t) dy}{K(\bold{x})}\bigg).
\end{align}
We will call this the ``canonical" model, as later on we will also be interested in a general family of such models.

%\begin{align}
%\label{modi1}
% \frac{\partial N(\bold{x}, t)}{\partial t} =  N(\bold{x}, t)\Bigg( 1 - \bigg(\frac{\int \alpha(\bold{x}, \bold{y}) N (\bold{y}, t) dy}{K(\bold{x})}\bigg)^m\Bigg).
%\end{align}
%For $m =1$, we recover Eq.(\ref{logistic}) from Eq.(\ref{modi1}).

Here $N(\bold{x}, t)$ is the population density of the phenotype $\bold{x}$ at time $t$ and $K(\bold{x})$ is the carrying capacity of the population consisting entirely of $\bold{x}$ individuals. 
The competition between individuals of phenotypes $\bold{x}$ and $\bold{y}$ is given by the competition kernel $\alpha(\bold{x}, \bold{y})$ and an individual of phenotype $\bold{x}$ faces competition with an effective density $\int \alpha(\bold{x}, \bold{y}) N (\bold{y}, t) dy$. We set the intrinsic growth parameter $r$ and $\alpha(\bold{x},\bold{x})$ to be equal to $1$. We consider $\bold{x}$ and $\bold{y}$ to be vectors of dimension $d$ describing a multidimensional phenotypic space.

The canonical equations of adaptive dynamics can be derived from the stochastic individual based model in the limit of rare mutations, small mutational effects and infinite population sizes. 
Under these assumptions, Dieckmann and Law showed that adaptive dynamics is the first order approximation of the mean path averaged over infinitely many realizations of the stochastic simulations obtained from the individual based model. In order to derive the adaptive dynamics of trait $\bold{x}$, we assume a monomorphic (Dirac-delta distribution) resident population in trait $\bold{x}$  with a globally stable equilibrium density given by $K(\bold{x})$ independent of the dimension $d$ of $\bold{x}$. A key ingredient to obtain the ODEs describing the adaptive dynamics is the \textit{invasion fitness} function whose gradient determines the direction of selection forces and hence of the deterministic evolutionary trajectory.
The \textit{invasion fitness} function is the difference between the per capita birth and per capita death rates of a rare mutant in a monomorphic population. 
The \textit{invasion fitness}, for the logistic map, of a rare mutant $\bold{y}$ is its per capital rate of growth in a resident population with phenotype $\bold{x}$ and is
given by
 \begin{align}
 \label{fitness}
f(\bold{x}, \bold{y}) = 1 - \frac{ \alpha(\bold{x}, \bold{y}) K(\bold{x})}{K(\bold{y})}.
\end{align}
In the above example for the logistic equation, the selection due to frequency dependence influences the death rates while the birth rates are neutral. In general, birth rates might also depend on interactions between individuals and the fitness function will have a more general form involving competition kernels for both birth and death.

From the \textit{invasion fitness}, we can derive the selection gradient to be
 \begin{align}
\label{sg1} 
s_i(\bold{x}) = \frac{\partial f(\bold{x}, \bold{y})}{\partial y_i}_{\big | \bold{y}=\bold{x}} =  - \frac{\partial \alpha(\bold{x}, \bold{y})}{\partial y_i}_{\big | \bold{y}=\bold{x}} + \frac{\partial K(\bold{x})}{\partial x_i}\frac{1}{K(\bold{x})}.
\end{align}
Finally, the canonical equation of adaptive dynamics in terms of the selection gradient is given by
 \begin{align}
 \label{AD1}
 \frac{d\bold{x}}{dt} = M(\bold{x}). s(\bold{x}).
\end{align}
Equation (\ref{AD1}) is a system of $d$ dimensional ODEs that describe a trajectory that may converge to a fixed point, a limit cycle, a quasi periodic orbit or exhibit chaos depending on the nature of the selection gradient $s(x)$ and the initial conditions.
 Here $M(x)$ describes the mutational process and for simplicity we assume it to be a $d \times d$ identity matrix. 

  Individual based models describe the interactions between discrete individuals that possess multi-dimensional adaptive traits.
  Such a microscopic description is more general and in a sense more fundamental as compared to its deterministic macroscopic approximation like the adaptive dynamics which can be derived from it under suitable assumptions.
 In order to simulate individual based models, we adapt the Gillespie algorithm \cite{Gillespie1976} for simulating chemical reactions to population biological stochastic processes.  Individuals are treated as particles, and birth and death events as chemical reactions. In the deterministic logistic model, the birth rate of phenotype $i$, $\r_{i}$, with trait value $\bold{x_i}$ is $1$. When a birth event happens, a new individual is added to the population with a phenotype that is offset from the parent by a small mutation, which happens with probability $\mu$, chosen from a uniform distribution with a small amplitude. 
 Throughout this paper we choose $\mu=1.0$. 
  Its death rate, ${\d_{i}}$ is given by $N_{eff}(i)/K(\bold{x_i})$, where $N_{eff}$ is the effective density experienced by phenotype $\bold{x}$. For the individual based model, if there are $N$ individuals, $\bold{x}_1$, ..., $\bold{x}_N$, at time $t$ then the effective density experienced by individual $i$ is
   \begin{align}
 \label{Neff} 
 N_{eff}(i) = \sum_{j \neq i} \alpha({\bold{x}_j, \bold{x}_i}),
\end{align}
and, therefore, the death rate is given by
\begin{align}
 \label{Neff}
 N_{eff}(i) = \sum_{j \neq i} N_{eff} /K(\bold{x}_i).
\end{align}

The individual based simulations are then implemented stochastically in the familiar way by performing one birth or death event \cite{Gillespie1976} that constitutes one computational step that advances the system from time $t$ to $t + \Delta t$. In the above example for the logistic equation, the selection due to frequency dependence influences the death rates while the birth rates are neutral. In general, both the birth as well as the death rates can be affected by frequency dependent interactions among individuals described by competition terms like $\alpha_{birth}({\bold{x}_j, \bold{x}_i)}$ and $\alpha_{death}({\bold{x}_j, \bold{x}_i)}$.

\section{A modified algorithm for individual based simulations}

As mentioned above, individual-based realizations of the model
are based on the Gillespie algorithm \cite{Gillespie1976} as described above and consists of assigning each individual $i$  a 
constant reproduction rate $\r_{i}=1$ and a death rate $\d_{i}=\sum_{\j\neq \i}
\alpha(\bx_{i},\bx{_j})/K(\bx_{i})$, as defined by logistic ecological dynamics.
 The total rate is
given by the sum of all individual rates $U=\sum_{i}
(\r_{i}+\d_{i})$. A particular ``event", birth or death, is chosen 
at random with probability equal to the rate of this event divided by the total probability
rate $U$. If a birth event is chosen, a new individual is added to the population with a phenotype that is offset from the parent by a small mutation chosen from a uniform distribution with a small amplitude.
Modified Gillespie algorithm consists of simply choosing the individual with the maximum death rate
and eliminating it when a death event is chosen. That is, we select $max(\delta_1, \delta_2, ..., \delta_n)$ and eliminate that individual. If both the birth as well as the death rates are affected by frequency dependent interactions among individuals described by competition terms like $\alpha_{birth}({\bold{x}_j, \bold{x}_i)}$ and $\alpha_{death}({\bold{x}_j, \bold{x}_i)}$, when the birth event is chosen, we select the individual with the maximum birth rate, $\r_{i}$, and a new individual is added to the population with a phenotype that is offset from the parent by a small mutation chosen from a uniform distribution with a small amplitude. We treat the death events analogously.

 It turns out that this simple modification has remarkable consequences as far as simulations are concerned. 
 We now discuss the consequences of this modification on the efficiency of the algorithm
and also on the adaptive dynamics it generates.

\subsection{Analysis of the Algorithm}
In order to understand the above algorithm and the reason behind its speed, we first consider a logistic model of the form

\begin{align}
\label{modilo}
 \frac{\partial N(\bold{x}, t)}{\partial t} =  N(\bold{x}, t)\Bigg( 1 - \bigg(\frac{\int \alpha(\bold{x}, \bold{y}) N (\bold{y}, t) dy}{K(\bold{x})}\bigg)^m\Bigg).
%%\label{mod_log}
\end{align}
The above model reduces to Eq. \ref{logistic} for $m=1$.
Assuming the resident to be at the equilibrium population density, given by the Dirac delta function,
$N(x') = K(x)\delta(x-x')$. Then the per capita growth of a mutant $y$ is given by

\begin{align}
\label{pcmod}
 \frac{\partial N(\bold{y}, t)}{\partial t N(\bold{y}, t)} =\Bigg( 1 - \bigg(\frac{\int \alpha(\bold{x'}, \bold{y}) K(x) \delta(x-x') dx'}{K(\bold{y})}\bigg)^m\Bigg).
\end{align}
Therefore, the invasion fitness, as defined above is given by
\begin{align}
\label{IFmod}
f(\bold{x}, \bold{y}) =\Bigg( 1 - \bigg(\frac{ \alpha(\bold{x}, \bold{y}) K(x) }{K(\bold{y})}\bigg)^m\Bigg).
\end{align}
From this, we can derive the selection gradient to be
\begin{widetext}
 \begin{align}
\label{msg1}
s_i(\bold{x}) = \frac{\partial f(\bold{x}, \bold{y})}{\partial y_i}_{\big | \bold{y}=\bold{x}} =  m \bigg(\frac{ \alpha(\bold{x}, \bold{y}) K(x) }{K(\bold{y})}\bigg)_{\big | \bold{y}=\bold{x}} \bigg(-\frac{\partial \alpha(\bold{x}, \bold{y})}{\partial y_i}_{\big | \bold{y}=\bold{x}} + \frac{\partial K(\bold{x})}{\partial x_i}\frac{1}{K(\bold{x})}\bigg), 
\end{align}
\end{widetext}
which simplifies to
\begin{widetext}
\label{msg}
\begin{align}
s_i(\bold{x}) = \frac{\partial f(\bold{x}, \bold{y})}{\partial y_i}_{\big |\bold{y}=\bold{x}} = - m\bigg(\frac{\partial \alpha(\bold{x}, \bold{y})}{\partial y_i}_{\big | \bold{y}=\bold{x}}+\frac{\partial K(\bold{x})}{\partial x_i}\frac{1}{K(\bold{x})}\bigg).
\end{align}
\end{widetext}
Therefore, the above selection gradient is $m$ times the selection gradient given in Eq. \ref{sg1}.

Finally, the canonical equation of adaptive dynamics in terms of the selection gradient is given by
 \begin{align}
 \label{AD2}
 \frac{d\bold{x}}{dt} = m M(\bold{x}). s(\bold{x}),
\end{align}
which gives the same set of ODEs up to the scaling constant $m$ as obtained previously with the canonical logistic model. We also note that the Eq.\ref{AD2} evolves $m$ times faster than Eq. \ref{AD1}.

In constructing the individual based models for the logistic model in Eq. \ref{modilo}, 
we note that the birth rates remain constant  at $\r_{i}=1$, while the  death rate for the individual $i$ is given by $\d_{i}=\big(\sum_{\j\neq \i}
\alpha(\bx_{i},\bx{_j})/K(\bx_{i})\big)^m$.
Therefore, the vector of these death rates can be expressed as  $(\delta_1^m, \delta_2^m, ..., \delta_n^m)$
where $(\delta_1, \delta_2, ..., \delta_n)$ are the death rates of individuals corresponding to the ``canonical" map. Let $k$ be the individual with the highest death rate. This means
$max(\delta_1^m, \delta_2^m, ..., \delta_n^m) = \delta_k^m$.
When $m$ is sufficiently large, we have $\delta_k^m >>\delta_i^m$ for all $i \neq k$. Therefore the probability of choosing $k$ to die is close to 1. But this is exactly what we do in our modified algorithm!
We deterministically pick the individual with the highest death rate and eliminate it. Therefore, the modified algorithm can be considered to be an individual based simulation of a birth death process governed by the logistic model given by Eq. \ref{modilo}. The value of $m$ in Eq. \ref{modilo} should be sufficiently large 
such that, we satisfy, $\delta_k^m >>\delta_i^m$ for all $i \neq k$.
We have already shown that the individual based simulations based on this model, in the deterministic limit, give rise to exactly the same equations of adaptive dynamics as the 
``canonical" model up to the scaling constant $m$.   

\subsection{Conditions for Diversification}
We have shown that the individual based simulations based on the modified algorithm gives us the same adaptive dynamics equations as the standard algorithm. ODEs describing the adaptive dynamics are first order differential equations describing the evolution of a monomorphic population in the phenotypic space.
In order to study diversification, we need to explore higher orders of the fitness function. 

Adaptive diversification has been extensively studied at singular points of the dynamics that also happen to be evolutionarily unstable \cite{mdbook, Geritz98, Metz92}. The canonical equation of adaptive dynamics has the form
\begin{align}
 \label{CAD}
 \frac{dx_i}{dt} = g_i(x),
\end{align}
    where the functions $g_i$ are given as
    \begin{align}
 \label{AD}
 \begin{pmatrix} g_i(x) \\ ... \\ g_d(x)\end{pmatrix} = \begin{pmatrix}  \frac{\partial f(x, y)}{\partial y_i}_{\big | y=x}  \\ ... \\  \frac{\partial f(x, y)}{\partial y_d}_{\big | y=x} \end{pmatrix}.
\end{align}
Singular points of the dynamics are the points, $\bold{x}^{*}$ at which the R.H.S. of Eq. (\ref{CAD}) is zero. 
%Singular point, $x^{*}$ is a repellor for the dynamics when the time evolution leads all points sufficently close to $x^{*}$ to diverge from it. This happens when the Jacobian at the singular point $J(x^*) >0$.
Singular point, $\bold{x}^{*}$ is called \textit{convergence stable} if the dynamics starting at all points sufficiently close to it converge to $\bold{x}^{*}$ eventually. This occurs when the Jacobian at the singular point, $J(\bold{x}^*) <0$.
These points are of great importance in adaptive dynamics as they can be potentially the points at which diversification occurs. 

For scalar traits, this can be seen by Taylor expanding the invasion fitness function with respect to the mutant $y$ to the second order, we get
\begin{widetext}
\begin{align}
 \label{TEF}
 f(x,y) = f(x, x) + \frac {\partial f(x, y)}{\partial{y}}_{\big |y=x}.(y-x)+  \frac{\partial^2 f(x, y)}{\partial y^2}_{\big | y=x}.\frac{(y-x)^2}{2}.
\end{align}
\end{widetext}
The first term on the RHS, $f(x, x)$ is zero for all $x$.
Usually, the evolutionary dynamics can be accurately described by the first order term in (\ref{TEF}) or, in other words, by the canonical equation of adaptive dynamics and we do not need to consider higher orders terms.
It is only in the neighborhood of singular points, we have $ \frac {\partial f(x, y)}{\partial{y}}_{\big | y=x} \rightarrow 0$, the second order term in Eq. (\ref{TEF}) becomes significant. In particular, at the singular point we have $ \frac {\partial f(x, y)}{\partial{y}}_{\big | y=x} = 0$ and if $ \frac{\partial^2 f(x, y)}{\partial y^2}_{\big | y=x} < 0$, no nearby mutants can invade the resident population that is monomorphic and we have the conditions for evolutionary stability. In contrast,  $ \frac{\partial^2 f(x, y)}{\partial y^2}_{\big | y=x} > 0$ is the condition for evolutionary instability, or potential evolutionary branching points, as the mutant now can potentially invade the resident.

It is easy to see that the fitness functions derived from Eq. \ref{logistic} and Eq. \ref{modilo} yield the same conditions for diversification. For the Eq. \ref{logistic} we have

\begin{align}
\label{div_1}
 \frac{\partial^2 f(x, y)}{\partial y^2}_{\big | y=x}   = -\frac{\partial}{\partial y}\Bigg(\frac{\partial}{\partial y}\bigg(\frac{\alpha(x, y) k(x)}{k(y)}\bigg)\Bigg)_{\big | y=x},
\end{align}

and for the invasion fitness function in Eq. \ref{IFmod}, the second derivative is
\begin{widetext}
\begin{align}
 \frac{\partial^2 f(x, y)}{\partial y^2}_{\big | y=x}   =-m\bigg(\frac{\partial}{\partial y}\bigg(\frac{\alpha(x, y) k(x)}{k(y)}\bigg)\bigg)^2_{\big | y=x} -m\frac{\partial}{\partial y}\Bigg(\frac{\partial}{\partial y}\bigg(\frac{\alpha(x, y) k(x)}{k(y)}\bigg)\Bigg)_{\big | y=x}.
\end{align}
\end{widetext}
At equilibrium, when we have $ \frac {\partial f(x, y)}{\partial{y}}_{\big | y=x} = 0$, the first term on the RHS is zero. And the second term coincides with Eq. \ref{div_1}. Therefore, the conditions for potential diversification are identical for both maps. Hence we conclude that to study diversification through the modified algorithm, we can expect to get similar results as we would using the standard Gillespie algorithm.

\section{Interpretation of the Canonical Equation}
 Canonical equation can be viewed as dynamics that unravel as individuals go through the birth-death process described above. We will give a geometric interpretation of the canonical equation through the birth- death process.
 
 We first consider the time reversal of the dynamics in the logistic model.
  We first recall that a birth-death process in the logistic model has birth and death rates given by $\r_{i}=1$ and $\delta_{i}=\sum_{\j\neq \i}
\alpha(\bx_{i},\bx{_j})/K(\bx_{i})$. If we simply interchange the birth rates with the death rates, i.e.,  $\d_{i}=1$ and $\r_{i}=\sum_{\j\neq \i}
\alpha(\bx_{i},\bx{_j})/K(\bx_{i})$, this gives us the time reversal of the evolutionary dynamics given by the logistic model. This can also be seen from the expression for the invasion fitness, Eq. \ref{fitness}
where the first term on the RHS represents the per capita birth rate and the second term represents the per capita death rate \cite{dieckmann96}. Interchanging birth and death rates merely changes the sign of the fitness function and the selection gradient thereby reversing the evolutionary dynamics given in Eq. \ref{AD1}.
Therefore, a birth event in the forward time is equivalent to a death event in the time reversed picture and, likewise, a death event in the forward time is equivalent to a birth event in the time reversed picture.
 
When a birth event is chosen, a new individual is added to the population with a phenotype that is offset from the parent by a small mutation chosen from a uniform distribution with a small amplitude. Therefore, one can view the birth event as the displacement of the population cloud by a very small amount.
 In the time reversed picture, this birth event becomes a death event and will lead to the displacement of the population cloud in the opposite direction.
 
Considering the logistic model, we consider a time period $ \Delta \tau$ which is very small compared to the total time in which the dynamics unravels but nonetheless large enough that we have a large number of birth and death events occurring. For infinite population sizes, we will have a large number of birth-death events happening even when $\Delta \tau$ becomes very small. We also assume, the number of birth events $N_b$ will be equal to the number of death events $N_d$ in the time period $\Delta \tau$. This is because the population is assumed to be at equilibrium and, the probability of birth is equal to the probability of death event, i.e $P(Birth) = P(Death)$. Therefore, invoking the law of larger numbers, the sequence of birth-death events will be typical sequences with $N(birth) = N(death)$.
Consider a monomorphic population cloud at a particular point $\bold{x}$ in the phenotypic space. In the limit of very small mutations and a sufficiently small $ \Delta \tau$, a single birth event will cause a displacement of the population cloud given by $\Delta r_i$. Since for the logistic model, the birth events are neutral, and therefore $\sum_i{\Delta r_i} = 0$.  The displacement by a single death event is, $\Delta q$, and the direction of displacement is in the opposite direction to the dying individual  $k$, where  $ \delta_k =max(\delta_1, \delta_2, ..., \delta_n)$.
$\d_{k}=max\big(\sum_{\j\neq \i}
\alpha(\bx_{i},\bx{_j})/K(\bx_{i})\big))$. Assuming all individuals except $k$ to be a monomorphic population, $\d_{k}=max\big(
\alpha(\bx,\bx +\Delta q)/K(\bx +\Delta q)\big))$. Thus, we seek to determine a vector displacement $\Delta q$ that will maximize $\d_{k}$. If $\Delta q$ is in the direction of the gradient of the function $ \frac{ \alpha(\bold{x}, \bold{y}) K(\bold{x})}{K(\bold{y})}$ with respect to $y$ at the value $x$. We assume the same direction for $\Delta q$ for all the death events occurring in the time interval $\Delta \tau$.
Therefore, the total net displacement is given by
 \begin{align}
 \label{netdisplacement}
 \sum_i{\Delta r_i} + \sum_j{\Delta q_j} = N_{death}\nabla_{\bold{y} |\bold{y} = \bold{x}} f(\bold{x}, \bold{y}). |\Delta q|.
\end{align}
 Taking the limits, $N_{death} \rightarrow \infty$, $|\Delta q| \rightarrow 0$ and $\tau \rightarrow 0$
 we get the canonical equation of adaptive dynamics, 
 \begin{align}
 \label{AD_basic_der}
 \frac{d{\bold{x}}}{dt} = \gamma \nabla_{\bold{y} |\bold{y} = \bold{x}} f(\bold{x}, \bold{y}).
\end{align}
 
 In general and as mentioned above, birth rates might also depend on interactions between individuals and the fitness function will have a more general form involving competition kernels for both birth and death.  Let, $\alpha_{birth}({\bold{x}_j, \bold{x}_i)}$ and $\alpha_{death}({\bold{x}_j, \bold{x}_i)}$ be the competition kernels for birth and death respectively. Since we assume an infinite population limit at equilibrium we have
 $N(birth) = N(death)$. In this case, we will need to add up the displacement by both birth as well as death events. For a single birth event, the displacement  is, $\Delta q_{b}$, and the direction of displacement is in the direction of the newly born individual  $k$, where  $ \r_k =max(\r_1, \r_2, ..., \r_n)$.
$\r_{k}=max\big(\sum_{\j\neq \i}
\alpha_{birth}(\bx_{i},\bx{_j})/K(\bx_{i})\big))$. Assuming all individuals, except $k$, as a monomorphic population, $\r_{k}=max\big(
\alpha_{birth}(\bx,\bx +\Delta q_b)/K(\bx +\Delta q_b)\big))$. Thus, we seek to determine a vector displacement $\Delta q_b$ that will maximize $\r_{k}$. If $\Delta q_b$ is in the direction of the gradient of the function $ \frac{ \alpha_{birth}(\bold{x}, \bold{y}) K(\bold{x})}{K(\bold{y})}$ with respect to $y$ at the value $x$. We assume the same direction for $\Delta q_b$ for all the birth events occurring in the time interval $\Delta \tau$. The calculation for the net displacement, $\Delta q_d$, due to death events follows similarly. Therefore, the total net displacement, with $N(birth) = N(death)$, is given by
  
 \begin{align}
 \label{netdisplacement1}
 \sum_i{\Delta r_i} + \sum_j{\Delta q_j} = N_{death}\nabla_{\bold{y} |\bold{y} = \bold{x}} f'(\bold{x}, \bold{y}). |\Delta q|,
\end{align}
where, $ f'(\bold{x}, \bold{y}) = \Big( \frac{ \alpha_{birth}(\bold{x}, \bold{y}) K(x) }{K(\bold{y})} - \frac{ \alpha_{death}(\bold{x}, \bold{y}) K(x) }{K(\bold{y})}\Big)$ is the per capita growth rate of a mutant $y$ in a resident population. In other words, $ f'(\bold{x}, \bold{y})$ is the invasion fitness function and we recover canonical equation of adaptive dynamics.
    
We interpret the above equation as the translation of a population cloud along the direction of instantaneous gradient of the fitness function with respect to the mutant phenotype. It is remarkable that in this derivation, unlike the standard derivation \cite{dieckmann96} which essentially considers the evolutionary trajectory as a sequence of invasions by the mutant phenotypes,
 we do not need to explicitly invoke concepts like ``invasion fitness" but recover it through the translation of the population cloud undergoing a birth-death process. 
 
Equation \ref{netdisplacement1} also suggests a faster way to implement birth death process when we have competition kernels for both birth and death events. At each iteration, we simply choose exactly one birth and one death event. For the birth event, the individual with the maximum birth rate is chosen and a new individual in its vicinity is produced. For the death event, as before, the individual with the maximum death rate is chosen to die. It is clear from the form of the invasion fitness function, Eq. \ref{fitness}, consisting of two separate terms for birth and death that such an algorithm will satisfy the canonical equation of adaptive dynamics.
The advantage we get is faster simulations and also the populations in the stochastic simulations does not increase and remains manageable. This is in contrast to the standard algorithm, where the population typically increases after evolutionary branching and the simulations computationally intensive after some time.
   
\section{Examples}

In this section, we illustrate the efficiency of the modified algorithm with the help of numerical examples.
We will also address the issue of unpredictability in evolution. 

\subsection{Evolution of a single trait}
Our first example compares the performance of  the standard and modified Gillespie algorithm for a single trait.
For this, we use the ``canonical" logistic model given in Eq. \ref{logistic}. We consider a Gaussian
competition kernel given by
 \begin{align}
 \label{GauCK}
 \alpha(x, y) = \exp\big[-\frac{(x-y)^2}{2\sigma_{\alpha}^2}\big].
\end{align}
The magnitude of $\sigma_{\alpha}$ tells us about the strength of the competition.

For the carrying capacity, we assume a Gaussian form as well,
\begin{align}
 \label{GauCC}
 K(x) = K_{0}\exp\big[-\frac{(x-x_{0})^2}{2\sigma_{k}^2}\big].
\end{align}
$K_{0}$ refers to the maximum carrying capacity and $x_{0}$ is the phenotype with the maximum carrying capacity. The condition for diversification \cite{mdbook}, $ \frac{\partial^2 f(x, y)}{\partial y^2}_{\big | y=x} > 0$, becomes
\begin{align}
 \label{GauCC}
\sigma_{\alpha} < \sigma_{k}.
\end{align}
This can be interpreted as a competition between the relative effect of the competition and carrying capacity
at the singular point. There has been great progress theoretically as well as a large amount of emperical evidence for symatric speciation \cite{gavrilets, barluenga06, BerlocherandFeder02, CoyneandOrr04, Débarre14, DoebeliDieckmann00, DoebeliIspolatov10, Friesen004, Gilman12, HerronandDoebeli13, ItoDieckmann07, ItoandDieckmann14, Le2012, Leimar2009, Leimar13, Plucain14, RaineyandTravisano98, Rosenzweig94, RozenandLenski00, Ryan07, Savolainen06, Svardal14, imd15}.
Fig 1 gives an example of the comparison between the standard implementation of the stochastic birth-death process and our modified procedure. The details of parameter values for the individual based simulations is given in the caption. We see the modified algorithm is far more efficient than the standard Gillespie algorithm. While the latter takes roughly $8 \times 10 ^4$ iterations for the first evolutionary branching to occur, the former shows multiple evolutionary branching events over $2 \times 10 ^4$ iterations.

\begin{figure*}
[t]\resizebox{17cm}{!}
{\includegraphics{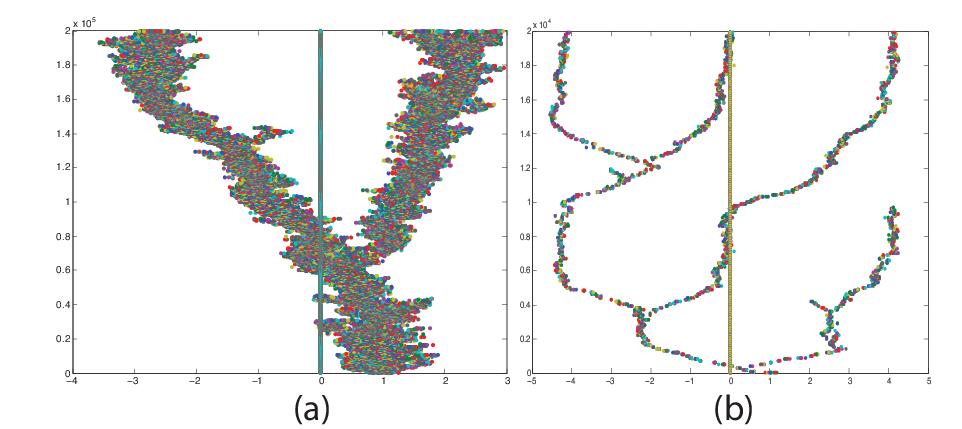}}
\caption{Side by side comparison of the performance of simulation of single trait adaptive dynamics over the number of iterations on the computer (vertical axis) using the standard Gillespie algorithm on the left (Fig 1(a)) and the modified algorithm on the right (Fig 1(b)). Parameter values for both (a) and (b): $\sigma_k = 2.0$, a Gaussian competition kernel with $\sigma_{\alpha} = 0.75$. $K_{0} = 200$. Initial population of 200 individuals was initialized about $x=1.0$ with a small variance. The mutation size after the birth event has a magnitude $\sigma_{\mu}=0.1$ and occurs with a probability $\mu=1.0$. We see the modified algorithm is far more efficient than the standard Gillespie algorithm. While the latter takes roughly $8 \times 10 ^4$ iterations for the first evolutionary branching to occur, the former shows multiple evolutionary branching events over $2 \times 10 ^4$ iterations.}
\label{F1}
\end{figure*}

\subsection{Evolutionary chaotic dynamics in a multi dimensional phenotypic space and the question of predictability of evolution}

 We assume that the complexity of the epistatic interactions
between phenotypic components is determined by the competition
kernel $\alpha(x, y) : \mathbb{R}^{d} \times \mathbb{R}^{d} \rightarrow \mathbb{R}$, which is, in general, a complicated non-linear function.
In particular, when we have the competition kernel of the ``canonical" form \cite{imd15},
\begin{align}
 \label{CCC}
\alpha({\bold{x}, \bold{y}}) = \exp(\sum_{j=1}^{d} b_{ij}x_{j}(x_i - y_i) + \sum_{j,k=1}^{d} a_{ijk} x_jx_k(x_i - y_i)),
\end{align}
 we get%consider only the Taylor expansion of this function up to quadratic terms, we get
  \begin{align}
 \label{sg}
 \frac{\partial \alpha(\bold{x}, \bold{y})}{\partial y_i}_{\big | y=x} = -\sum_{j=1}^d b_{ij} x_j - \sum_{j,k=1}^d
a_{ijk}x_j x_k.
\end{align}
 For the carrying capacity function, we assume it to be of the form $K(x) = \exp (-\sum_i x_i^4/4)$ .
Then (\ref{AD1}) can be written as 
\begin{align}
\label{main1}
\frac{dx_i}{dt}=\sum_{j=1}^d b_{ij} x_j + \sum_{j,k=1}^d
a_{ijk}x_j x_k - x_i^3, \; i=1,\ldots,d.
\end{align} 
 The above set of ODEs describe the adaptive dynamics of the monomorphic population in the phenotypic space. In \cite{id14, ispolatov15}, it was shown that the dynamical system described by Eq. (\ref{main1}) exhibits chaos with almost unit probability for sufficiently high dimensional phenotypic space, i.e. $d \geq 50$. %\cite{id14}.  
 
 In particular, we are interested when (\ref{AD1}) results in chaotic behavior.
 	The above model can also be transformed to give the adaptive dynamics that coincides with the known chaotic systems
like the Lorenz model. We just need to fix the coefficients
$a_i$, $b_j$ and $c_k$ of the competition kernel in the above model. For example, specific choice of coefficients  will lead to Lorenz model and other popular models in the literature to
explore the stochastic effects inherent in the birth death simulations when the corresponding equations for the adaptive dynamics are chaotic.
 Assuming the adaptive dynamics to be governed by the equations of the Lorenz model, we then proceed to study and construct the corresponding individual based model and simulate it the modified version of the standard algorithm.

\begin{figure*}
[t]\resizebox{14cm}{!}
{\includegraphics{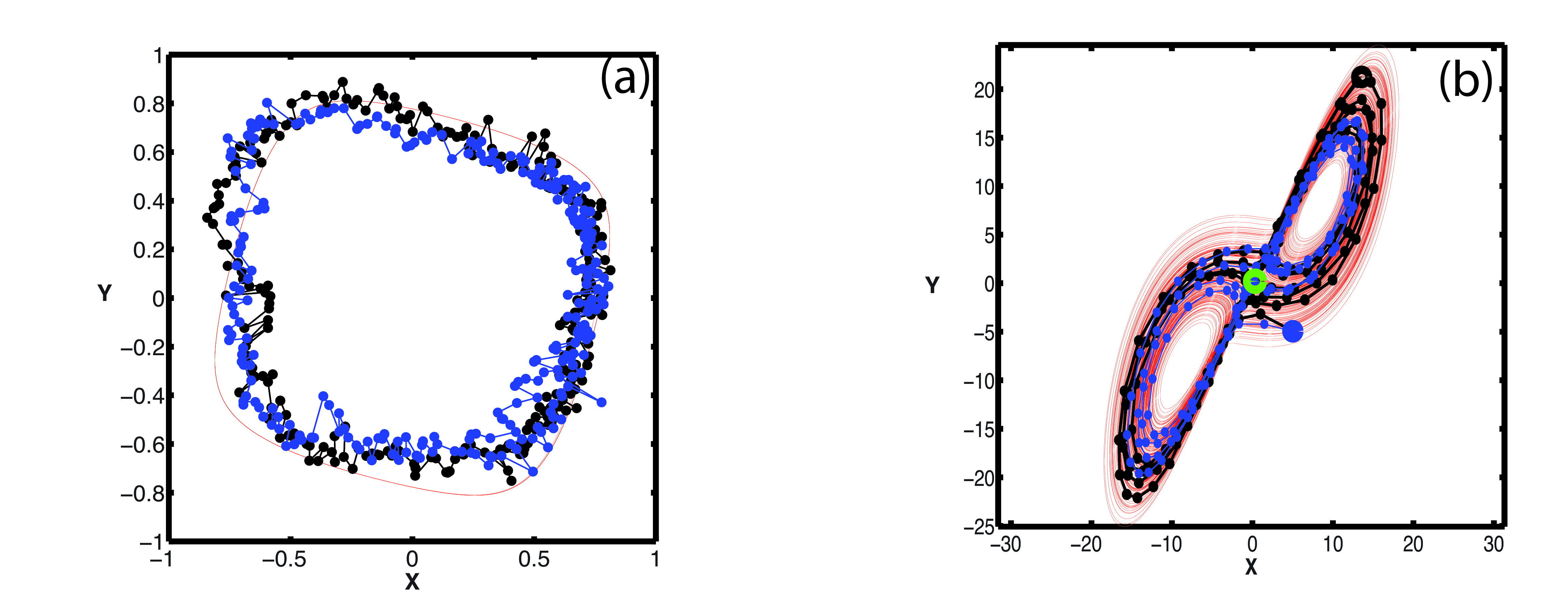}}
\caption{Examples of correspondence between the monomorphic adaptive dynamics and the individual based model with the canonical competition kernel. (a) When the adaptive dynamics exhibits a limit cycle (red). Individual based models (black and blue) follow the monomorphic trajectory with stochastic fluctuations due to a small population size (about 100 individuals). (b) When the adaptive dynamics is the Lorenz attractor (red). Individual based models (black and blue) show the ``butterfly effect". Starting from \textit{exactly same} initial conditions (the green dot), they evolve to different locations on the attractor (the black and the blue dot). (c) Example of another chaotic adaptive dynamics (red) and correspondence with the individual based models (black and blue).}
\label{F2}
\end{figure*}

Starting two simulations from exactly the same initial conditions and same mutation rates we 
 see that the evolutionary trajectories follow completely different paths after sometime \ref{F2}.
Therefore the intuition about the evolutionary trajectories occupying two completely different regions of phenotypic space is correct. And this occurs purely due to a combination of stochastic effects due to mutations and the underlying chaotic adaptive dynamics.
Firstly, this tells us, contrary to the traditional view, that evolution is not a directional optimization process that goes to equilibration \cite{id14}. It might be a chaotic trajectory that is ergodic and sensitive to the initial conditions as well as the to the nature of very small mutational changes for exactly the same initial conditions.
 Even more fascinating is the sensitive dependence of a chaotic evolutionary trajectory on very small mutations, or the ``butterfly effect" which casts a serious doubt over our ability to predict evolution over long time scales. 
 
 Fig 2 demonstrates the ``evolutionary butterfly effect" for two instances of  individual based simulations starting from exactly the same initial conditions
 when the underlying adaptive dynamics is chaotic. This is contrasted with individual based simulations of a cyclic adaptive dynamics which remains predictable throughout its course. 
%Last but not the least, from an ecological stand point, it is crucial to investigate the the long term fate of multiple population clusters coexisting in multi-dimensional phenotypic space with issues relating to the saturation of diversity and adaptive radiation.

%\section{Analysis of the Modified Algorithm}

%\begin{figure*}
%\centering
%[t]\resizebox{18.1cm}{!}
%\includegraphics[width=20cm]{diversification_1.pdf}
%\caption{Examples of diversification through a competition kernel with the addition of a Gaussian component. Distinct population clusters representing speciation are shown as black clouds.(a) When the adaptive dynamics exhibits a limit cycle. There is permanent diversification in this case. (b) When the adaptive dynamics is the Lorenz attractor. Each black ``dot" represents a population cloud. (c) Example of a four dimensional chaotic adaptive dynamics and emergence of diversity.}
%\label{F2}
%\end{figure*}  
%\begin{widetext}
%\begin{align}
 %\label{GCCC}
%\alpha_g({\bold{x}, \bold{y}}) = \exp\Big(\sum_i \frac{(x_i - y_i)^2}{2 \sigma_i^2}\Big)\exp\Big(\sum_{i=1}^{d} b_{ij}x_{j}(x_i - y_i) + \sum_{j,k=1}^{d} a_{ijk} x_jx_k(x_i - y_i)\Big),
%\end{align}
%\end{widetext}
  %will yield the same first order adaptive dynamics given by Eq. (\ref{main1}) above.
%\clearpage

%\section{Examples}

%\section{The modified algorithm as a limit of the regular algorithm}

\section{Conclusion} 
Individual based models are essential in order to study various aspects of evolution like diversification,
predictability, adaptive radiation and frequency dependence. It has been observed that these models
follow the adaptive dynamics trajectories remarkably well, even when the underlying adaptive dynamics is 
chaotic \cite{imd15}. In this work, we derive the canonical equation of adaptive dynamics in a single step 
from a modified algorithm for simulating individual based models. We therefore provide a very simple and intuitive way of understanding the correspondence between adaptive dynamics and individual based models.
Using a simple modification to the Gillespie algorithm, we were able to simulate individual based birth-death models with significant efficiency. With the help of examples, we demonstrate the efficiency of  the modified algorithm.  We also demonstrate how the modified algorithm can be interpreted as simulating the individual based model associated with a generalized logistic equation as given by Eq. \ref{modilo} and how this yields exactly the same equations for adaptive dynamics as the standard logistic model. We also discuss how the conditions for evolutionary branching for both algorithms remain same. Lastly, we discussed how to simulate the birth death process when both birth and death events have a competition kernel associated with them while, at the same time, keeping the total population size bounded despite of evolutionary branching. Our analysis gives an intuitive interpretation of the canonical equation of adaptive dynamics as the translation of the population cloud in the phenotypic space due to birth-death events. Such an approach has interesting connections to the stochastic quantization in quantum field theory and the birth death events can be viewed as creation and annhilation operators of a quantum field \cite{bianconi}. A complete analysis of these connections is the focus of our future work.

Our simulations also enables us to address important questions regarding the nature of evolutionary dynamics.
  In \cite{id14} it was shown that phenotypic properties can combine in complicated ways and as a result evolutionary trajectories can exhibit chaos. It shows that, contrary to the traditional view, evolution is not a directional optimization process that goes to equilibration. Evolution might be a complicated trajectory in the phenotypic space that is continuously twisting, turning and folding upon itself on a chaotic attractor.
 Our numerical simulations involving individual based simulations confirm that chaotic trajectories of evolution are sensitive to very small mutational changes. Even when these trajectories start with exactly the same initial conditions, they can end up in completely different regions of the phenotypic space. Therefore, if Gould's tape of life were to be run again, with exactly same initial conditions and dynamics, we can end up with completely different biodiversity on earth. Moreover, the size of the mutations to cause this unpredictability of evolution need not be big as the underlying chaotic dynamics amplify small errors arising due to mutations. We hope our work is useful for simulations involving birth death processes in high dimensional phenotypic spaces as well as the study of evolution of language, culture, religions and agent based models in economics.

\end{document}